# Thermal decoupling in high-$T_c$ cuprate superconductors

*Sungwoo Lee[1,2,3]*, Woojin Choi[1,2,3], Yeongjae Kim[2,3], Young-Kyun Kwon[4], Dongjoon Song[5], Miyoung Kim[2,3], and Gun-Do Lee[1,2,3]**

[1]Center for High-temperature Superconductors, Seoul National University, Seoul 08826, Republic of Korea

[2]Department of Materials Science and Engineering, Seoul National University, Seoul 08826, Republic of Korea

[3]Research Institute of Advanced Materials, Seoul National University, Seoul 08826, Republic of Korea

[4]Department of Physics, Department of Information Display, and Research Institute for Basic Sciences, Kyung Hee University, Seoul 02447, Republic of Korea

[5]Stewart Blusson Quantum Matter Institute, University of British Columbia, Vancouver, British Columbia V6T 1Z1, Canada

*corresponding authors: gdlee@snu.ac.kr, blessing1223@gmail.com

**Abstract**

In unconventional high-$T_c$ cuprate superconductors, the intricate interplay between the non-ergodic bad metal and the strange metal state has remained enigmatic. Herein, we unravel this mystery using ab initio molecular dynamics simulations and the temperature-dependent effective potential method. Our investigation, centered on $YBa_2Cu_3O_7$, provides the first simulation report on the $B_{1g}$ phonon anomaly, unveiling thermal decoupling induced by the bond weakening between the Ba atom and $CuO_2$ plane. This decoupling emerges as a pivotal underpinning behind several puzzling phenomena in high-$T_c$ superconductivity. Our results indicate that the effective temperature on the BaO plane deviates from that of the $CuO_2$ plane at low temperatures. Furthermore, we delineate the correlation between thermal decoupling and the Planckian dissipation, rigorously and quantitatively revealing a connection between linear-$T$ resistivity, Uemura relation, and superconducting domes, which are known to be the most important unsolved mysteries of high-$T_c$ superconductivity. The suppressed isotope effect ($\boldsymbol{\alpha} \approx 0.02$) in cuprates is also quantitatively explained from thermal decoupling. Our discoveries offer a revolutionary perspective on high-$T_c$ superconductivity, suggesting the potential for a transformative shift in our comprehension. Furthermore, they suggest that the autonomous emergence of low-temperature layers within materials has the potential to revolutionize industrial thermal management challenges.

## 1. Introduction

The high-$T_c$ cuprate superconductors distinguish themselves from conventional superconductors in a profound way: their parent compounds are Mott insulators [1,2]. Through the process of doping, these insulators transition into a state termed as a bad metal, marked by suboptimal electron transport properties due to extremely short mean free paths [3,4]. Remarkably, it was suggested that the bad metal in a many-body localized state can manifest as a non-ergodic metal, a unique condition where the system fails to achieve thermodynamic equilibrium over extended durations [5]. The intricate relationship between this bad metal and the strange metal observed in the normal state of high-$T_c$ superconductors remains an enigma. Recent advancements have ushered in a renewed interest in the realm of superconductivity, exemplified by observations in magic-angle twisted bilayer graphene (MATBG) with flat bands [6,7]. The revelation that flat bands can induce a disorder-free many-body localization (MBL) has garnered attention [8,9]. MBL is characterized by the system failing to reach thermal equilibrium. Furthermore, the nexus between the MBL and the onset of superconductivity is a subject of fervent exploration among many researchers. A salient characteristic of the bad metallic state in cuprates is the marked augmentation in the specific heat coefficient ($\gamma$) of the electrons, especially at the optimal doping levels where the $T_c$ maximizes [10]. Given that the $\gamma$ is directly proportional to the electron effective mass [11], this phenomenon hints at diminished electron mobility, coupled with an augmented effective mass in the optimal doping regime. This leads to the intriguing proposition that, contrary to conventional metal, phonons, rather than electrons, primarily drive heat transfer in cuprates, which has been experimentally substantiated [12]. Hence, the present paper delves into these intriguing features of cuprate superconductors, aiming to elucidate the link between the bad metal and the mysterious strange metal. Motivated by these findings, we seek to elucidate the interplay of these thermal phenomena in cuprates, studying the phonon itself in relation to thermal properties rather than its role in mediating Cooper pairs in electron-phonon interactions. A phonon anomaly is a representative phonon phenomenon that appears peculiarly in high-$T_c$ cuprate superconductors [13-15]. Numerous studies have been conducted on phonon anomalies as they relate to superconductivity. In particular, the $B_{1g}$ phonon anomaly appears just above $T_c$ when the temperature is lowered, and the $B_{1g}$ phonon frequency drops rapidly due to softening below $T_c$ [13,16,17]. The $B_{1g}$ phonon anomaly is believed to originate from the depletion of the electron density of states due to the gap opening[18]; however, the relationship between the two has not been resolved. We conducted a comprehensive study on the phonon anomaly, considering a potential correlation with the unique thermal phenomena arising from the low electron mobility in cuprates.

In this study, we investigated the phonon anomaly in the representative cuprate, $YBa_2Cu_3O_7$ (YBCO) using the ab initio molecular dynamics (AIMD) simulation method and the temperature-dependent effective potential (TDEP) method [19,20]. YBCO is the most intriguing and innovative cuprate superconductor, and the material that sparked research into high-temperature cuprate superconductivity [21,22]. Notably, our study is the first simulation report on the $B_{1g}$ phonon anomaly. Beyond the confirmation of the $B_{1g}$ phonon anomaly in YBCO, we also revealed thermal decoupling inside YBCO that is induced due to the decoupling between the thermal properties and electron transport. Through rigorous analysis, we elucidate that this thermal decoupling plays a pivotal role in deciphering enigmatic phenomena in high-$T_c$ superconductivity, including the strange metal, the Uemura relation, the superconducting dome, and the suppressed isotope effect.

## 2. Methods

We performed AIMD simulations using the Vienna *ab initio* simulation package (VASP) code [23,24] with a Nosé–Hoover thermostat [25,26] controlling the temperature. Plane-wave pseudopotential calculations using ultrasoft pseudopotentials [27] were carried out within the framework of the generalized gradient approximation of the Perdew–Burke–Ernzerhof functional [28]. The energy cut-off for the plane-wave basis was set to 400 eV. The DFT-D3 scheme proposed by Grimme et al. was used to approximate the long-range dispersion forces [29]. In this simulation, we adopted a 4×4×1 bulk supercell and one Γ point for k-point sampling in the Brillouin zone. We confirmed that the supercell and k-point sampling are sufficient to describe the phonon anomaly in YBCO accurately. All AIMD simulations were performed with a time step of 1 femtosecond within the *NVT* ensemble. After we first performed an AIMD simulation at 300 K for over 60 picoseconds (ps), we continued AIMD simulations at each temperature for 60 ps by decreasing the temperature by 10 K for 5 ps from 300 to 20 K. In this study, we tested the magnetism of YBCO by performing spin-polarized AIMD simulations for 5 ps at 300 K, 90 K, and 30 K, and obtained almost non-magnetic results with total magnetic moments smaller than 0.0005 $\mu_B$ per 1×1×1 unit cell at all AIMD simulation steps. The thermal decoupling data at 90 K and 30 K also yielded results very similar to those of the spin-unpolarized calculations. Our study also shows that a spin-unpolarized calculation successfully describes the B1g phonon anomaly, demonstrating that it is sufficient to capture phonon-induced thermal decoupling accurately in this system. In unconventional high-$T_c$ superconductivity research, spin fluctuations may be important from an electronic perspective, but our research focuses on phonon-centered thermal phenomena, and we have confirmed that considering spin has little effect on thermal phenomena. Therefore, we performed full AIMD simulations within a spin-unpolarized scheme to save huge computational time. To obtain the phonon properties of YBCO, the effective force constant matrices were calculated using the TDEP code with atomic displacements and forces from the AIMD at finite temperatures [19,20]; this process was carried out by least-squares fitting of AIMD displacements and forces to the crystal model Hamiltonian. The TDEP method can extract the force constant from the AIMD results and is useful for temperature-dependent phonon studies including anharmonic effects [30]. We rigorously tested the convergence of the phonon properties of YBCO in the TDEP calculations. (For the detailed simulation structure of YBCO, see Discussion S1 in the supplemental material (SM))

## 3. Results

### 3.1. $B_{1g}$ phonon anomaly and $A_g$ phonon anomaly from Ba atoms

We performed AIMD simulations at various temperatures, decreasing from 300 K to 20 K, and obtained temperature-dependent phonon dispersions using the TDEP method. Fig. 1(a) shows the frequency of the $B_{1g}$ phonon as a function of temperature in YBCO. The $B_{1g}$ phonon mode is the out-of-plane vibration mode of oxygen atoms in the $CuO_2$ plane. Our simulation shows that the frequency of the $B_{1g}$ phonon increases as the temperature decreases from 300 to 150 K, with a more gradual decrease around 150 K, similar to the behavior observed in experiments [13,16,17]. These results demonstrate that our simulation methods accurately reproduce the phonon-induced dynamics observed in experiments on YBCO. However, the frequency drops abruptly near 70 K, which is slightly different from what was observed experimentally, in which the $B_{1g}$ phonon frequency drops abruptly near the superconducting temperature, at ~ 90 K. We will discuss this difference later in our analysis. Intriguingly, this anomalous behavior occurs not only in the $B_{1g}$ phonon but also in the $A_g$ phonon, which is even stronger, as shown in Fig. 1(b). The $A_g$ phonon is an out-of-plane mode mainly from Ba atoms. The frequency of the $A_g$ phonon decreases by ~ 70 cm$^{-1}$ from 300 to 20 K, whereas that of the $B_{1g}$ phonon decreases by just 1.3 cm$^{-1}$. Such an $A_g$ phonon anomaly originating from Ba atoms has been reported in Raman spectroscopy for YBCO [31]. However, no serious anomaly has been reported, which could be due to the thermalization associated with spectroscopy measurements. Phonons with low frequencies are generally affected by observation processes such as those involving lasers [32]. Notably, many other out-of-plane phonons, as well as the $B_{1g}$ and $A_g$ phonons, are softened when the temperature is reduced below $T_c$ (See Fig. S1 in the SM), which has also been observed in Raman spectroscopy [33].

### 3.2. Thermal decoupling from phonon analysis and AIMD simulations

To investigate the effect of phonon softening from the phonon anomalies in the thermodynamics of YBCO, we analyzed the phonon dispersion relation with respect to temperature. Fig. 2(a) shows phonon dispersions obtained from TDEP calculations carried out after AIMD simulations at 300, 90, and 30 K. The internal energy of a harmonic crystal is given by the following [34]:

$$U = Vu^{\text{eq}} + \sum_{\mathbf{k}s} \frac{1}{2}\hbar\omega_s(\mathbf{k}) + \sum_{\mathbf{k}s} \frac{\hbar\omega_s(\mathbf{k})}{e^{\beta\hbar\omega_s(\mathbf{k})}-1} \quad (1)$$

where $V$ is the crystal volume, $\omega_s(\mathbf{k})$ is the frequency of the phonon modes, and $\beta$ is $1/k_B T$. The first and second terms represent the energy of the equilibrium configuration and the energy of zero-point vibration, respectively. The last term is the temperature-dependent part of the internal energy (TDPIE), of which the frequency dependence is displayed as the background color of Fig. 2(a). The contribution of Ba atoms for each phonon mode and the wave vector is also calculated from the phonon eigenstates, which is shown as the grayscale lines overlaying the phonon dispersion curves in Fig. 2(a). The contribution of Ba atoms for each phonon mode is distributed mainly in the low-frequency phonon modes; this contribution does not change significantly as the temperature decreases. This implies that the contribution ratio of Ba atoms to TDPIE increases as the contribution of high-frequency phonons to TDPIE decreases with the temperature. Figure 2(b) shows the contribution of each type of atom to TDPIE as a function of temperature. The contribution of Ba atoms to TDPIE increases as the temperature decreases, reaching a maximum near 30 K. Thus, the contribution of each type of atom to the internal energy is different at low temperatures. This behavior is expected to affect the thermal properties of YBCO, potentially generating local thermal phenomena at low temperature, depending on the type of atom. It is well known that free electrons in metals play an important role in thermal transport and are capable of delocalizing local thermal phenomena. Figure 2(c) shows the charge density distribution in YBCO, which reveals that the Ba atom is separated electronically from the other atoms, especially from the Cu and O atoms in the adjacent $CuO_2$ plane. This result indicates that the thermal properties in YBCO due to the specific thermal phenomena that emerged from Ba atoms can be separated from the electronic properties determined by the $CuO_2$ plane. It has been reported that the non-equilibrium steady state is sustained and the Wiedemann–Franz law could break down when heat and particle flow are decoupled or weakly coupled at low temperatures [35]. The deep softening of the out-of-plane phonons from Ba atoms and O atoms of the $CuO_2$ plane due to the $B_{1g}$ and $A_g$ phonon anomalies indicates bond weakening between the two that sharply reduces the thermal transport, resulting in the thermal decoupling between the Ba atom and $CuO_2$ plane.

To explore further the thermal decoupling in YBCO, we calculated the kinetic energy from the velocity of atoms at each time step of the AIMD simulation and extracted layer-by-layer effective temperature ($T_{\mathrm{eff}}$) from the kinetic energy-temperature relation ($\frac{1}{2}mv^2 = \frac{3}{2}k_{\mathrm{B}}T_{\mathrm{eff}}$). Figure 3(a)-(c) show the $T_{\mathrm{eff}}$s of the BaO plane, $CuO_2$ plane, Y layer, and CuO chain at the simulation temperatures of 300, 90, and 30 K, respectively. At 300 K, almost all layers have the same $T_{\mathrm{eff}}$, whereas, at the simulation temperatures of 90 and 30 K, the $T_{\text{eff}}$ of the BaO plane is higher than those of the other planes. Our results indicate that the high $T_{\text{eff}}$ of the BaO plane is mainly due to the Ba atoms, as shown in Fig. 3(e) and (f). This is consistent with our analysis of the temperature-dependent phonon dispersion shown in Fig. 2(b).

## 3.3. Analysis of unconventional superconductor mysteries via thermal decoupling

### 3.3.1. Linear-T resistivity

The question is then whether this thermal decoupling can be observed experimentally. It has been reported that the surface of YBCO is stabilized mainly with the termination of the BaO plane [36]. The stability of the BaO surface from our AIMD simulation is also discussed (See Fig. S2 in the SM). In the AIMD simulation, the BaO surface acts as a protective surface layer for YBCO. Therefore, the measured temperature in experiments is likely to coincide with the temperature of the BaO surface. We attempted to scale the system temperature to the $T_{\text{eff}}$ of the BaO plane to re-analyze the $B_{1g}$ phonon anomaly curve shown in Fig. 1(a). The $B_{1g}$ phonon anomaly depending on $T_{\text{eff}}$ of the BaO plane is shown in Fig. 4(a); notably, the experimental observations exhibited an abrupt drop in the $B_{1g}$ phonon curve near the superconducting temperature (~ 90 K). These results not only show that our simulation accurately reproduces the experimental results for the phonon anomaly, but also demonstrate that the measured temperature in the experiment could be the $T_{\text{eff}}$ of the BaO plane. We attempted to clarify the relationship between the $T_{\text{eff}}$ of the surface BaO plane ($T_{\text{BaO}}$) and the $T_{\text{eff}}$ of the conducting $CuO_2$ plane ($T_{\text{CuO}_2}$) from our AIMD simulation in Fig. 4(b). From the optimally fitted curve for the normal state at $T_{\text{BaO}} >$ 100 K, as shown in Fig. 4(b), we found that the two temperatures are coupled to $T_{\text{BaO}} = 2.08 \times 10^{-3}\, {T_{\text{CuO}_2}}^{2.02} + 108.2$. Because the fitting equation satisfies essentially the quadratic relation of $T_{\text{BaO}} \propto {T_{\text{CuO}_2}}^{2}$, we refitted their relationship quadratically to obtain $T_{\text{BaO}} = 2.31 \times 10^{-3}\, {T_{\text{CuO}_2}}^{2} + 107.6$.

We defined the decoupling temperature, $T_{\mathrm{dec}}$, below which the thermal decoupling occurs between the two planes, as the temperature at which the slope of the quadratic fitting curve becomes one, implying that $T_{\mathrm{BaO}}$ and $T_{\mathrm{CuO_2}}$ approach the same value. This value was estimated to be $T_{\mathrm{dec}} = 216.3$ K from our AIMD simulation results, as shown in Fig. 4(b). Here, we demonstrate that this thermal decoupling is the origin of the famous phenomenological formula in high-$T_c$ cuprate superconductivity, i.e. the Planckian dissipation [37,38]:

$$\frac{\hbar}{\tau} = \alpha k_{\mathrm{B}} T \tag{2}$$

where $\tau$, $\alpha$, and $T$ are the relaxation time, a dimensionless parameter, and the measured temperature, respectively, and $\hbar$ and $k_{\mathrm{B}}$ are Planck's and Boltzmann's constant, respectively. The Planckian dissipation expresses the linear-$T$ resistivity behavior in the strange metal phase of the cuprates. However, the origin of the Planckian dissipation has not been understood until now. $\tau$ should be determined by the conducting $CuO_2$ plane. Since the stable surface of YBCO is the BaO plane, the measured temperature $T$ in experiments can be regarded as $T_{\mathrm{BaO}}$. By replacing $T$ by the quadratic fitting formula in the Planckian dissipation [Eq. (2)], we obtained $\hbar/\tau \propto 2.31 \times 10^{-3}\, {T_{\mathrm{CuO_2}}}^2$, indicating that the Planckian dissipation can be expressed like a Fermi liquid, as a quadratic term of $T_{\mathrm{CuO_2}}$. Therefore, linear-$T$ resistivity is due to thermal decoupling, which results from the temperature measurement of surface phonons in the BaO plane. This is illustrated in Figure 5.

### 3.3.2. Uemura relation

In this section, we quantitatively explain the Uemura relation, one of major mysteries in unconventional high-$T_c$ superconductors, which shows the linear relationship between $T_c$ and $T_F$, through the thermal decoupling. As discussed in the previous section, since the $CuO_2$ plane of YBCO is expressed as Fermi liquid, the resistivity form of the $CuO_2$ plane is given as follows.

$$\rho_{\mathrm{Fermi}} = A\frac{k_{\mathrm{B}}{T_{\mathrm{CuO_2}}}^2}{\hbar T_F}\frac{m^*}{ne^2} \tag{3}$$

where $A$ is a dimensionless parameter from the Fermi liquid; and $T_F$, $m^*$, and $n$ are the Fermi temperature, the effective mass, and the carrier concentration, respectively. However, below $T_{\mathrm{dec}}$, the measured temperature is expressed as $T_{\mathrm{BaO}}$ and the resistivity will follow Planckian dissipation of Eq. (2) (See Fig. 5):

$$\rho_{\mathrm{Planckian}} = \alpha\frac{k_{\mathrm{B}}T_{\mathrm{BaO}}}{\hbar}\frac{m^*}{ne^2} \tag{4}$$

Here, we consider the superconducting critical temperature of the $CuO_2$ plane as $T_0$. The superconducting temperature ($T_{\mathrm{c}}$) for the BaO plane will differ, depending on $T_{\mathrm{dec}}$, as shown in Fig.5 and Fig. S3 in the SM. The resistivity and the derivatives of Eq. (3) and Eq. (4) should be the same at $T_{\mathrm{dec}}$. For the relation, we obtain the relation between $T_{\mathrm{dec}}$ and $T_F$ as follows.

$$T_{\mathrm{dec}} = \frac{\alpha}{2A}T_F \tag{5}$$

The superconducting critical temperature, $T_{\mathrm{c}}$, is also derived as follows (see Discussion S2 in the SM for the detailed derivation):

$$T_{\mathrm{c}} = \frac{A}{\alpha}\frac{{T_0}^2}{T_F} + \frac{\alpha}{4A}T_F \tag{6}$$

$$\approx \frac{\alpha}{4A}T_F \tag{7}$$

where the approximation comes from $T_F \gg T_0$. This result explains the linear relationship between $T_{\mathrm{c}}$ and $T_F$ from the Uemura plot [39-41]. We compare the linear coefficient to corroborate more clearly the linear relation from the Uemura plot. Given that $T_{\mathrm{dec}}$ was calculated to be 216.3 K from our simulation results and $T_F$ is about 2450 K for $YBa_2Cu_3O_7$ [39], we estimate $\frac{\alpha}{2A}$ = 0.088 from Eq. (5). From Eq.

(7), we found $T_c/T_F$ = 0.044, which is in excellent agreement with $T_c/T_F$ = 0.042 for cuprates from the Uemura plot (See Discussion S3 in the SM) [41]. Therefore, Uemura relation is directly and quantitatively derived from our thermal decoupling. We also estimated $A$ = 5.67 with $\alpha \approx 1$ for almost all cuprates [11]. The coefficient $A$, which is the parameter from Fermi liquid resistivity in Eq. (3), is generally taken to be of an order unity within one power of ten [34], indicating that the resistivity of the $CuO_2$ plane is in good agreement with that of the general Fermi liquid. Discussion S4 in the SM provides a more detailed explanation of the relationship between $T_c$, $T_F$, and $T_{\mathrm{dec}}$ through thermal decoupling, and describes the resulting Uemura relation. In our study, we found that the two representative mysteries in unconventional high-$T_c$ superconductivity, specifically, the linear-$T$ resistivity and the Uemura relation, are closely connected and are explained by the thermal decoupling.

#### 3.3.3. Superconducting dome

We also revealed a relationship between the thermal decoupling and the superconducting temperature considering the distance between the Ba atom and $CuO_2$ plane. Even though, as mentioned, the Ba atom and $CuO_2$ plane are separated from each other electronically and thermally, a reduction in the distance between the two induces the thermalization of the $CuO_2$ plane, which affects $T_c$. The doping concentration in high-$T_c$ cuprate superconductors is a known key factor in determining $T_c$, which is called superconducting dome. However, their relationship has not been clearly resolved. The doping concentration affects the structure of YBCO and is determined by the number of oxygen atoms in the YBCO unit cell. As the number of oxygen atoms increases from $x = 6$ to 7 in $YBa_2Cu_3O_x$, the cell size $c$ along the $z$-direction becomes shorter. However, it is very intriguing that the Ba-$CuO_2$ plane distance becomes longer whereas the cell size $c$ becomes shorter [42-44]. From our results, we can conjecture that the change in distance between the Ba and $CuO_2$ planes affects $T_c$. Surprisingly, Fig. 6(a) shows that the variation of $T_c$ is well-matched with the Ba-$CuO_2$ plane distance modulated by the doping concentration.

Therefore, we need to consider superconducting phenomena from the viewpoint of thermal decoupling. In cuprates replaced by Sr instead of Ba, Sr could play the same role as Ba, as it is also an alkaline earth metal on the neighboring layer of the $CuO_2$ plane in Sr-doped cuprates. In our study expanded to other cuprate superconductors containing Ba and Sr, we also found that the distance between Ba or Sr and the $CuO_2$ plane is closely related to the enhancement of $T_c$ [45-54]. Figure 6(b) shows the relationship between $T_c$ and the distance between Ba or Sr and the $CuO_2$ plane in various cuprates. Among cuprates having a similar composition, $T_c$ increases with the distance, implying an enhancement in the thermal decoupling. Especially, in the case of the Hg-Ba-Ca-Cu-O compound (HBCCO), $T_c$ is closely related to the distance between Ba and the $CuO_2$ plane, regardless of the trend in the composition ratio. In the study of HBCCO under high pressure, the importance of the Ba position was reported [55]. In Discussion S5 of the SM, we also explain the suppression of $T_c$ for overdoped cuprates from the relationship between $T_c$ and $T_F$ which is attributed to the emergence of a Fermi liquid due to the reduction of thermal decoupling.

### 3.3.4. Suppressed oxygen isotope effect

We try to understand the suppressed isotope effect in high-$T_c$ cuprate superconductors from our thermal decoupling. First, we extract the superconducting temperature of $CuO_2$ plane, $T_0$ for underdoped cuprate superconductors where the thermal decoupling is well applied from the least-square fitting of experimental data for Eq. (6) (Discussion S6 in the SM). We found that $T_0$ is about 14 K (See also Discussion S6 in the SM) which is in excellent agreement with the lowest $T_c$ reported in YBCO [56]. It is intriguing that this value is within the limit of BCS theory. We demonstrate that the suppressed isotope effect can be quantitatively explained using Eq. (6) based on the result of $T_0 \approx 14$ K. The exponent $\boldsymbol{\alpha}$ value of the oxygen isotope effect in cuprate superconductors is experimentally known to be significantly suppressed compared to the BCS theory value of 0.5, averaging $\approx 0.02$ [57]. In Eq. (6), the temperature affected by the isotope effect is $T_0$ in the first term, as predicted by BCS theory. However, the overall $T_c$ is dominated by the second term of Eq. (6) when $T_F$ is large, and the second term is not affected by the isotope. Therefore, according to Eq. (6), the isotope effect is significantly suppressed. If the thermal decoupling effect decreases, the second term decreases, and as the $T_c$ decreases, the exponent $\boldsymbol{\alpha}$ approaches the BCS theory value of 0.5. Using Eq. (6), we calculated the isotope effect associated with replacing $^{16}O$ with $^{18}O$. For $YB_2C_3O_{6.6}$, the exponent $\boldsymbol{\alpha}$ is 0.016 as explained (Discussion S7 in the SM). Additionally, by calculating the average for twenty known cuprates with varying $T_c$ and $T_F$ values (Table SI and Table SII in the SM), we obtain 0.022 (Fig. 7 and Discussion S7 in the SM), which shows remarkable agreement with the experimental value ($\boldsymbol{\alpha} \approx 0.02$) [57]. Figure 7 also shows the agreement with experimental results showing that the exponent $\boldsymbol{\alpha}$ increases as the $T_c$ decreases [58]. Therefore, we propose that thermal decoupling can quantitatively explain the suppressed isotope effect in cuprate superconductors.

## 4. Discussion

The decoupling temperature ($T_{\text{dec}} = 216.3$ K) obtained in our simulation is similar to the temperature at which a kink in the thermal diffusivity of cuprates is observed [59] and the temperature at which the slope of linear-*T* resistivity changes [60]. In general, the linear-*T* resistivity in cuprates has been reported to persist up to higher temperatures. Recently, it has been reported that this linear-*T* resistivity is due to different mechanisms above and below about 200 K [3,59]. The linear-*T* resistivity at high temperatures is due to phonon scattering, as in conventional metals, but the linear-*T* resistivity at low temperatures below about 200K remains unclear. In this study, we suggested that thermal decoupling is the origin of linear-*T* resistivity at low temperatures. The elucidation of the mechanism of the connection between linear-*T* at high and low temperatures is beyond the scope of this study and requires further research.

We observed in Fig. 6 that the doping concentration is strongly correlated with thermal decoupling and directly influences $T_c$. Another factor significantly affecting $T_c$ is pressure. For high-pressure oxygenated YBCO, interesting results have been reported showing that $T_c$ does not decrease at all, even in highly overdoped regions [61]. Furthermore, it has recently been reported that $T_c$ remains almost unchanged even after pressure release [62-64]. The structure changes with pressure, and this could be related to thermal decoupling. Therefore, investigating the relationship between high pressure and thermal decoupling would also be intriguing.

Recent studies have been conducted on superconductivity in MATBG, as we mentioned in Section 1. The flat band discovered in MATBG is characterized by a small $T_F$. However, according to the Uemura relation observed in unconventional superconductivity up to now, superconductivity does not improve in flat bands with a small $T_F$. Therefore, to understand this contradictory case, research on quantum geometry is active [65-67], but the connection between flat bands and superconductivity has not yet been clarified. Eq. (6) shows that, for cuprates, the first term can be neglected, allowing the Uemura relation to be quantitatively explained. However, in the case of flat bands, the first term cannot be neglected, and $T_c$ increases when $T_F$ is small (See Fig. S4 in the SM). Therefore, our thermal decoupling suggests the possibility of enhanced superconductivity in MATBG. Recently, it has been reported that $T_c$ enhances beyond the Uemura relation when $T_F$ is extremely small [68]. Thus, our thermal decoupling suggests the possibility of extending the unconventional superconductivity to a flat band regime, such as MATBG, through Eq. (6). As mentioned in Section 1, in the case of flat bands, the effective mass of electrons becomes very large, slowing down the mobility of electrons, which is an essential condition for thermal decoupling. Therefore, studying thermal decoupling in flat bands will be an interesting topic in the future.

Our research findings indicate that $T_{\mathrm{BaO}}$ and $T_{\mathrm{CuO_2}}$ differ below $T_{\mathrm{dec}}$. Direct measurement of different temperatures in YBCO may be difficult with current technology. However, we propose two experiments that can indirectly confirm this temperature difference. First, in ARPES experiments, when the measured temperature is lowered at regular intervals, the FWHM near the Fermi level would decrease uniformly if all layers had the same temperature. However, if thermal decoupling occurs, the FWHM would show a different trend because it is determined by the electrons in the $CuO_2$ plane. Second, Johnson noise thermometry directly measures the temperature of the layer where electrons are actually transported. Therefore, when thermal decoupling occurs, the measured temperature representing BaO in YBCO will differ from the layer temperature at which electrons are transported.

From our results, heavy elements like Ba atoms act as a heat absorber due to the low-frequency phonons from those elements in phonon dispersion, which is important to the enhancement of $T_c$. The high-pressure-induced superconductivity from hydrides is a good example of the combination of heavy and light elements. It has been reported that the combination of heavy and light elements could generate MBL from quantum disentanglement [69]. MBL is a dynamic phenomenon observed in isolated many-body quantum systems that are characterized by non-equilibrium states. The decoupling between the thermal property and electron transport, and the resulting thermal decoupling in cuprates, can be treated as non-equilibrium thermodynamic states embodying the phase transition from volume law entanglement to area law entanglement at low temperature [70]. It would be interesting to study the relationship between this thermal decoupling and MBL. Bekenstein and Hawking also suggested that thermodynamics in the horizon of a black hole is dominated by area-law entanglement [71,72]. Recently, the connection between high-$T_c$ superconductivity and black hole physics has been studied using the Sachdev-Ye-Kitaev (SYK) model [73,74]. Therefore, it would also be interesting to study the SYK model from the perspective of thermal decoupling. Thermal decoupling may indicate the need to modify the measured temperature. In the experiment, careful speculation is also needed, as the temperature corresponding to critical physical properties may differ from the measured temperature. Above all, the most important result from our study is that a low-temperature layer relative to the surface layer can be generated spontaneously in solids, despite the lesser effort required for cryogenics. This may provide a unique means of solving frequently confronted thermal problems associated with quantum devices, quantum computing, and information transfer.

## 5. Summary

We successfully reproduced the $B_{1g}$ phonon anomaly in cuprate superconductors using the AIMD simulations and TDEP method, and discovered a severe anomaly in $A_g$ phonons caused by Ba atoms at low temperatures. We observed that thermal decoupling in YBCO arises from the softening of Ag phonons induced by Ba, and this thermal decoupling surprisingly explains the previously unsolved mysteries of linear-$T$ resistivity, the Uemura relation, and the superconducting dome at a quantitative level. In particular, we confirmed that linear-$T$ resistivity and the Uemura relation are closely connected through thermal decoupling. Additionally, the suppressed isotope effect observed in cuprate superconductors is also quantitatively explained. Therefore, we propose that thermal decoupling could serve as a powerful tool to quantitatively explain the experimental results of unconventional high-$T_c$ superconductors that have remained unexplained until now.

## CRediT authorship contribution statement

**Sungwoo Lee:** Writing – original draft, Methodology, Investigation, Formal analysis, Data curation. **Woojin Choi:** Investigation, Formal analysis, Data curation. **Yeongjae Kim:** Formal analysis, Data curation. **Young-Kyun Kwon:** Writing – review & editing, Investigation. **Dongjoon Song:** Writing – review & editing, Investigation. **Miyoung Kim:** Wring – review & editing. **Gun-Do Lee:** Writing – original draft, Wring – review & editing, Investigation, Formal analysis, Data curation, Conceptualization, Funding acquisition

## Declaration of competing interest

Authors declare no competing interests.

## Acknowledgements

The authors are grateful to Jan Zaanen, Paul C. W. Chu, Liangzi Deng, Zhifeng Ren, Keun Su Kim, and Philip Kim for helpful discussions. G.-D.L. acknowledges support from the Supercomputing Center/Korea Institute of Science and Technology Information with supercomputing resources (KSC-2022-CHA-0013 and KSC-2024-CHA-0015) and the National Research Foundation of Korea (NRF) grant funded by the Korean government (RIAM NRF-RS-2024-00416976). S.L. acknowledges financial support from the Korean government through NRF (NRF-2021R1I1A1A01050645). Y.-K.K. acknowledges financial support from the Korean government through NRF (NRF-2022R1A2C1005505 and NRF-2022M3F3A2A01073562). D.S. thanks in part to funding from the Canada First Research Excellence Fund, Quantum Materials and Future Technologies Program.

**Appendix A. Supplementary data**

Supplementary data to this article can be found online at https://doi. org/***.

**Data availability**

Data will be made available on request.

Figures

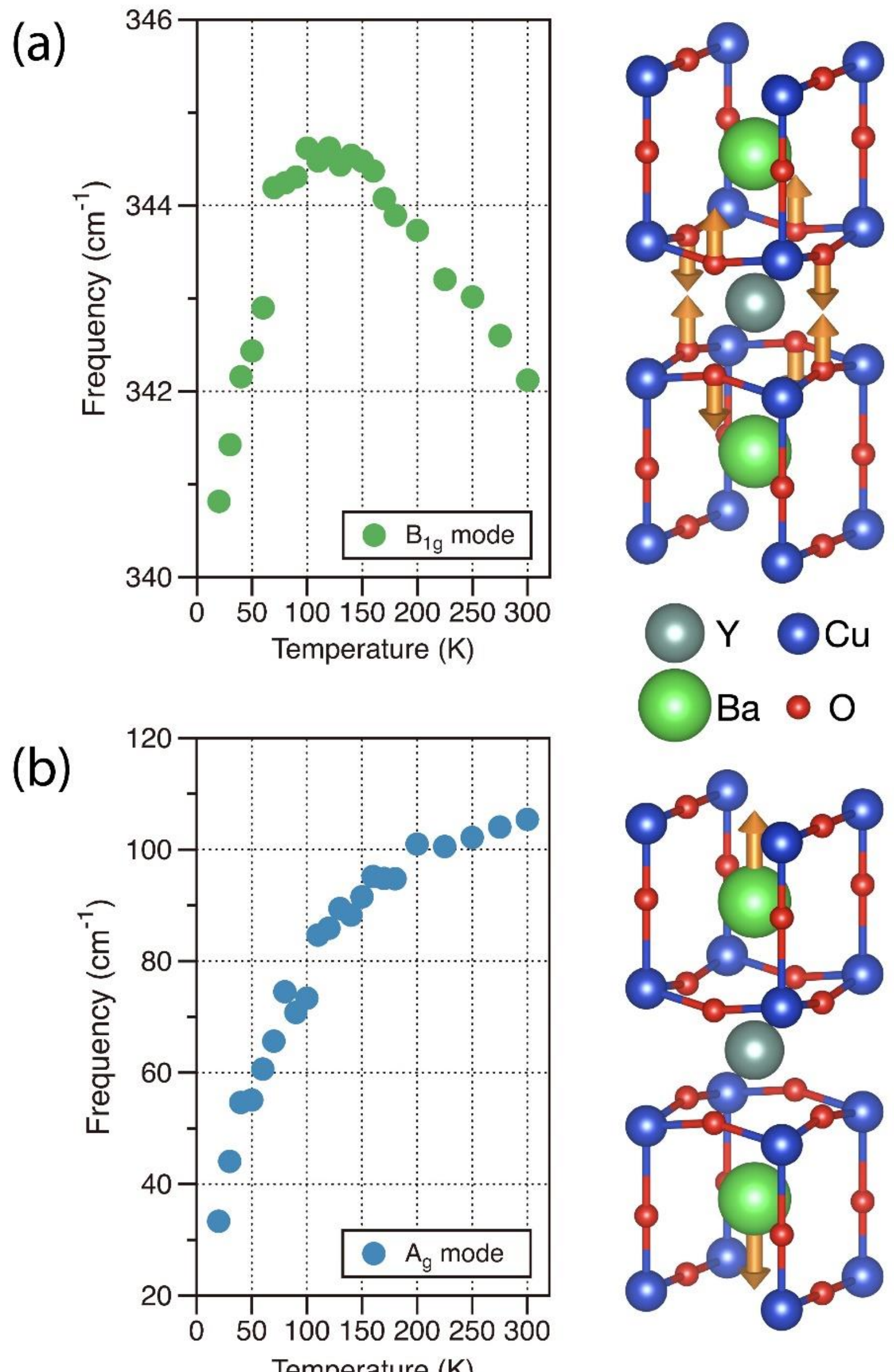

**Fig. 1. Phonon anomaly of the $B_{1g}$ and $A_g$ phonon modes.** (a) $B_{1g}$ phonon frequency dependence on temperature. (b) $A_g$ phonon frequency dependence on temperature (left), and their corresponding phonon modes visualized with the orange arrows in the real-space lattice (right)

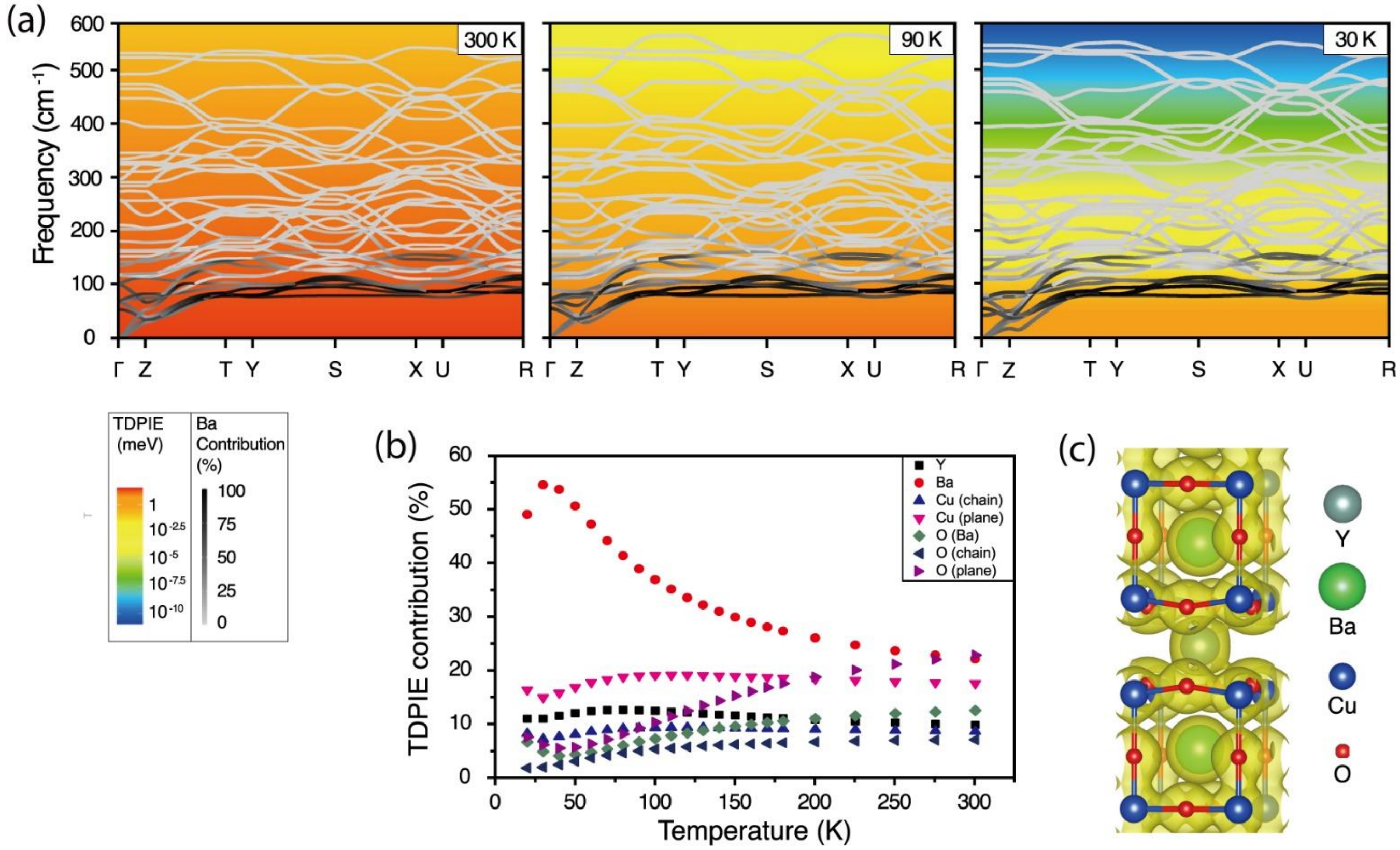

**Fig. 2. Thermal decoupling from phonon dispersion and charge density analysis.** (a) Phonon dispersions at 300, 90, and 30 K were calculated from TDEP. The colored backgrounds show the temperature-dependent part of internal energy (TDPIE) depending on the phonon frequency. The grayscale lines overlaying the phonon dispersion curves indicate the Ba contribution to each phonon mode evaluated by analyzing the phonon edge states. (b) Contribution of each type of atom to TDPIE. The contribution of Ba increases as the temperature decreases. (c) Charge density plot with the isosurface charge (0.047 electrons/$a_0^3$) exhibiting the electronic isolation of Ba atoms.

.

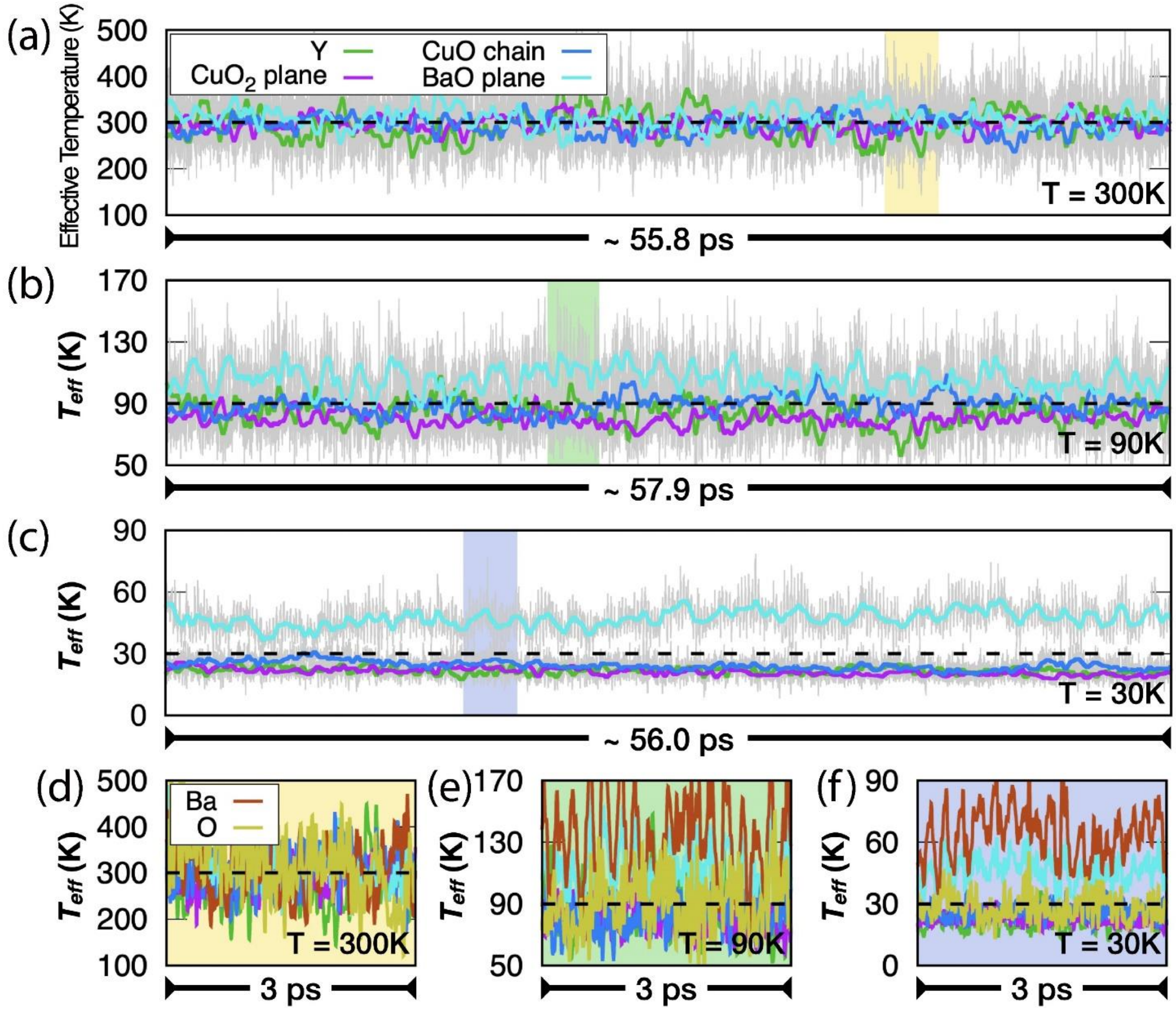

**Fig. 3. Thermal decoupling of YBCO at low temperature from *ab initio* molecular dynamics simulations.** (a-c) Layer-by-layer effective temperatures ($T_{\mathrm{eff}}$) at the system temperatures of 300, 90, and 30 K. (d-f) Magnified view of $T_{\mathrm{eff}}$ of the Ba and O atoms in the BaO layer in the yellow, green, and violet regions in (a), (b), and (c), respectively. In (a-c), the background grey peaks show the variation of real simulation temperatures, and the color curves show their Gaussian-broadened temperatures. In (d-f), the color curves show the real simulation temperatures.

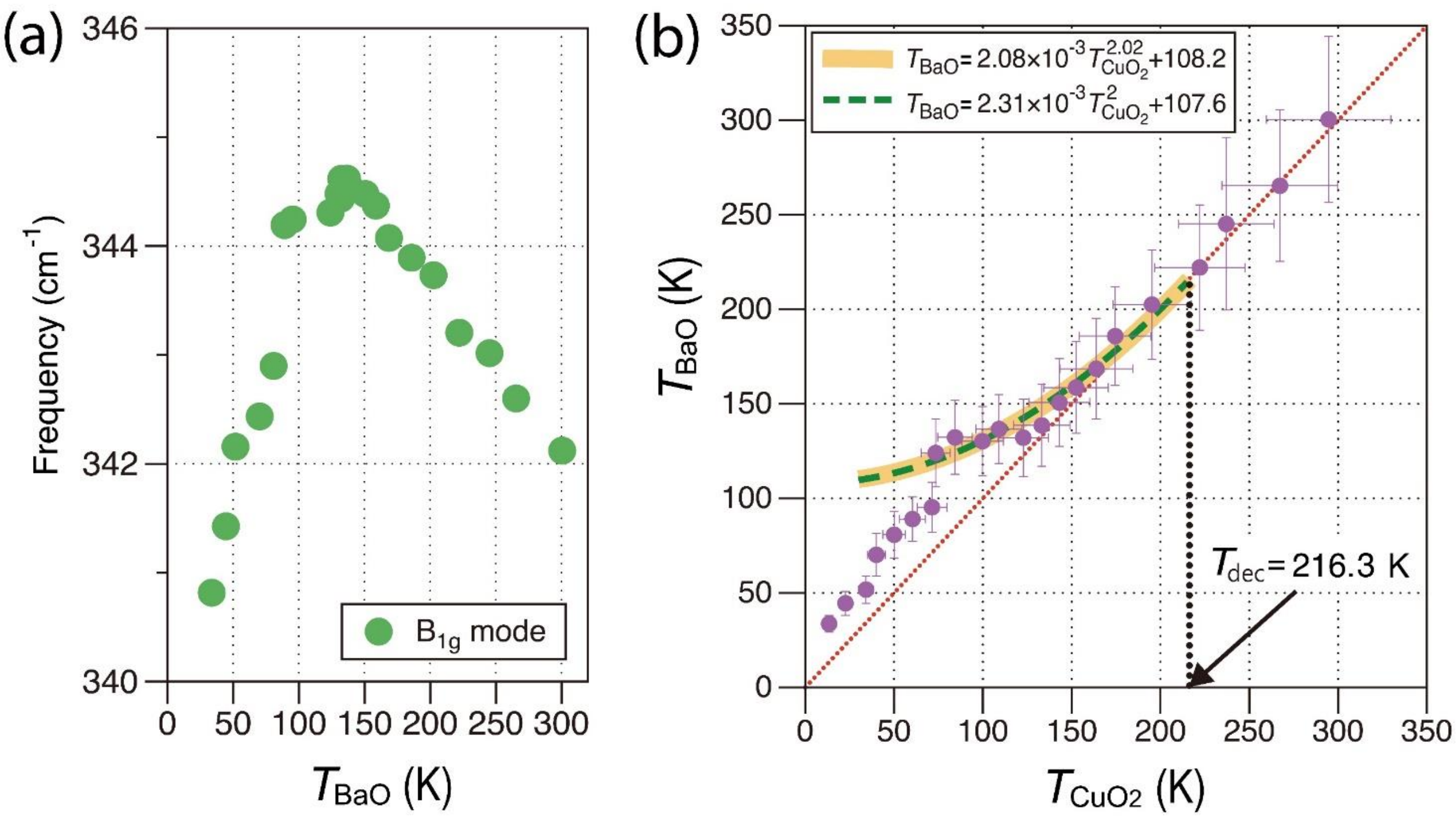

**Fig. 4. BaO-plane temperature ($T_{BaO}$) dependence of the $B_{1g}$ phonon and the relationship between $T_{CuO_2}$ and $T_{BaO}$.** (a) $B_{1g}$ phonon frequency as a function of $T_{BaO}$. It shows the shift in temperature compared to Fig. 1(a). (b) The relationship between $T_{CuO_2}$ and $T_{BaO}$. The purple dots are the average value of $T_{CuO_2}$ and $T_{BaO}$ from AIMD simulations and error bars represent the standard deviation of fluctuating effective temperatures with time. The fitting curves (thick yellow and dashed green lines) show the quadratic relation between $T_{BaO}$ and $T_{CuO_2}$. The arrow and the number indicate the decoupling temperature ($T_{dec}$) between the BaO and the $CuO_2$ planes

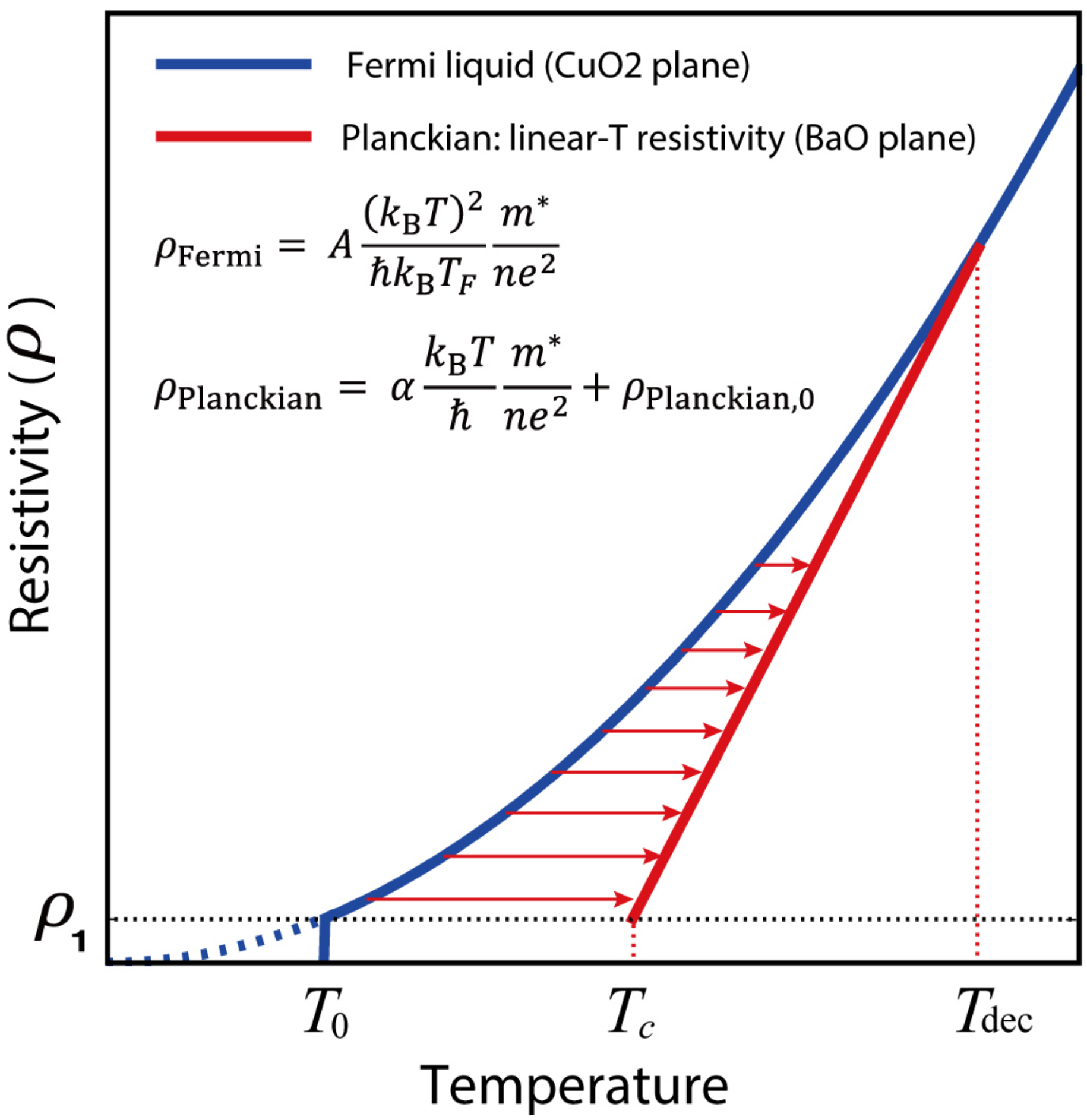

**Fig. 5. Resistivity relationship between the $CuO_2$ plane and the BaO plane.** The blue line shows the resistivity of the Fermi liquid for the $CuO_2$ plane. In terms of the BaO plane temperature ($T_{\mathrm{BaO}}$), the temperatures are shifted as indicated by the red arrows, and the resistivity becomes linear due to the relationship $T_{\mathrm{BaO}} \propto {T_{\mathrm{CuO_2}}}^2$ as explained in the text. The superconducting temperature of the $CuO_2$ plane is $T_0$, and the measured superconducting critical temperature becomes $T_{\mathrm{c}}$ from the BaO plane. $\rho_1$ is the resistivity of the $CuO_2$ plane just before the superconducting transition. $\rho_{\mathrm{Planckian,0}}$ is a constant that satisfies the resistivity relationship between the Fermi liquid and the Planckian dissipation

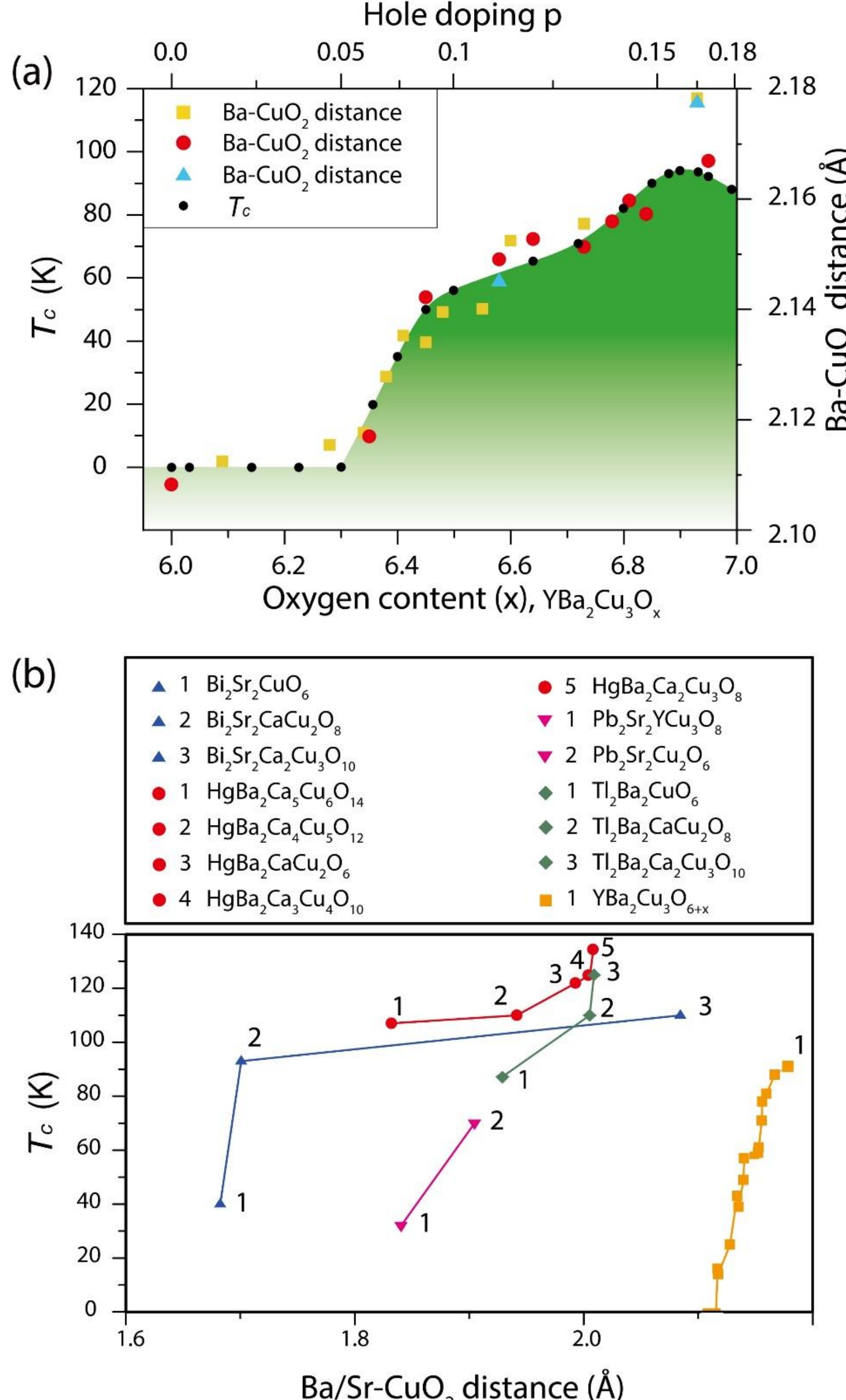

**Fig. 6. Relationship between $T_c$ and the Ba/Sr-CuO₂ plane distance in cuprates.** (a) Distance between the Ba atom and the $CuO_2$ plane on the doping concentration of YBCO is well matched with $T_c$. Data for the distance between the Ba atom and $CuO_2$ plane extracted from three experimental measurements were marked with yellow squares [42], red circles [43], and blue triangles [44], respectively. (b) The distance between the Ba or Sr atom and $CuO_2$ plane in various cuprate superconductors is matched with $T_c$ [45-54]

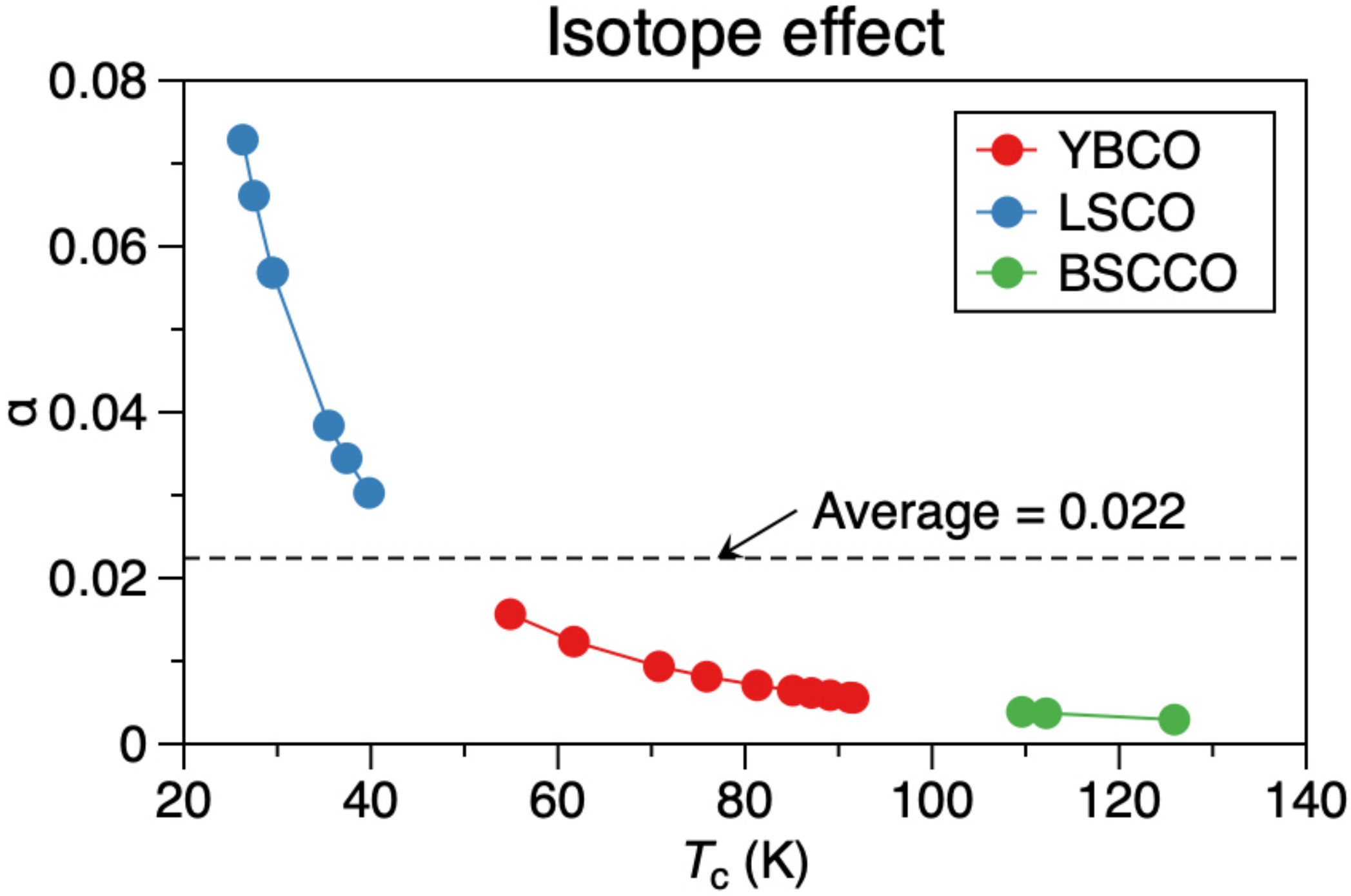

**Fig. 7. Exponents (α) in oxygen isotope effect of various cuprates calculated using Eq. (6)** (See Discussion S7 and Table SII in the SM). The dashed line represents the average of **α** data points

Supplemental Material for

# Thermal decoupling in high-$T_c$ cuprate superconductors

*Sungwoo Lee*[1,2,3]*, *Woojin Choi*[1,2,3], *Yeongjae Kim*[2,3], *Young-Kyun Kwon*[4], *Dongjoon Song*[5], *Miyoung Kim*[2,3], *and Gun-Do Lee*[1,2,3]*

[1]Center for High-temperature Superconductors, Seoul National University, Seoul 08826, Republic of Korea

[2]Department of Materials Science and Engineering, Seoul National University, Seoul 08826, Republic of Korea

[3]Research Institute of Advanced Materials, Seoul National University, Seoul 08826, Republic of Korea

[4]Department of Physics, Department of Information Display, and Research Institute for Basic Sciences, Kyung Hee University, Seoul 02447, Republic of Korea

[5]Stewart Blusson Quantum Matter Institute, University of British Columbia, Vancouver, British Columbia V6T 1Z1, Canada

*corresponding authors: gdlee@snu.ac.kr, blessing1223@gmail.com

**Supplemental Discussions**

**Discussion S1: Details of AIMD simulation structure of YBCO**

In AIMD simulation, we adopted the orthorhombic structure of $YBa_2Cu_3O_7$ (15.308 Å×15.652 Å×11.738 Å) that was extended by 0.08 Å along the *c*-axis compared to the *ab initio* optimized structure. In the simulation with the *ab initio*-optimized structure, we found that the phonon frequency was overestimated, which is related to a general tendency to overestimate the bulk modulus in *ab initio* calculations. Therefore, we performed TDEP calculations from AIMD results on the structure, in which the lattice constant was extended along the *c*-axis; the results were in good agreement with the experimental results of the $B_{1g}$ phonon frequency, as shown in Fig. 1(a). To check the robustness of our results, we also performed AIMD simulations and TDEP calculations for the *ab initio*-optimized structure and obtained phonon anomalies similar to those in Fig. 1 and layer-by-layer thermal decoupling (Fig. S6).

**Discussion S2: Derivation of Uemura relation from thermal decoupling**

The resistivity relationship between the $CuO_2$ plane and the BaO plane is explained in Fig. 5. The resistivity of the Fermi liquid for the temperature of the $CuO_2$ plane is given by

$$\rho_{\mathrm{Fermi}} = A\frac{(k_{\mathrm{B}}T)^2}{\hbar\varepsilon_F}\frac{m^*}{ne^2} \quad ,$$

as the form of Eq. (3).

The resistivity from Planckian dissipation for the temperature of the BaO plane is as follows:

$$\rho_{\mathrm{Planckian}} = \alpha\frac{k_{\mathrm{B}}T}{\hbar}\frac{m^*}{ne^2} + \rho_{\mathrm{Planckian,0}} \; . \qquad \text{(S1)}$$

To satisfy the resistivity relation between the Fermi liquid and the Planckian dissipation, we add the constant $\rho_{\text{Planckian,0}}$ to Eq. (4) as shown in Fig. 5. At $T_{\text{dec}}$, the slopes should be the same.

$$\left.\frac{d\rho_{\text{Fermi}}}{dT}\right|_{T_{\text{dec}}} = \left.\frac{d\rho_{\text{Planckian}}}{dT}\right|_{T_{\text{dec}}}$$

$$2A\frac{{k_{\text{B}}}^2}{\hbar\varepsilon_F}T_{\text{dec}} = \alpha\frac{k_{\text{B}}}{\hbar}.$$

Here,

$$\varepsilon_F = k_{\text{B}}T_F$$

Therefore,

$$T_{\text{dec}} = \frac{\alpha}{2A}T_F .$$

At $T_{\text{dec}}$, the resistivities should be the same.

$$\rho_{\text{Fermi}} = \rho_{\text{Planckian}} ,$$

$$A\frac{(k_{\text{B}}T_{\text{dec}})^2}{\hbar\varepsilon_F}\frac{m^*}{ne^2} = \alpha\frac{k_{\text{B}}T_{\text{dec}}}{\hbar}\frac{m^*}{ne^2} + \rho_{\text{Planckian,0}} ,$$

$$\rho_{\text{Planckian,0}} = A\frac{(k_{\text{B}}T_{\text{dec}})^2}{\hbar\varepsilon_F}\frac{m^*}{ne^2} - \alpha\frac{k_{\text{B}}T_{\text{dec}}}{\hbar}\frac{m^*}{ne^2} ,$$

If the $CuO_2$ plane becomes superconducting at $T_{\text{CuO}_2} = T_0$ and the resistivity just before the superconducting transition is $\rho_1$ (see Fig. 5), we can find $T_{\text{c}}$ from Eq. (S1) and $\rho_{\text{Planckian}} = \rho_1$:

$$T_{\text{c}} = \left(\rho_1 - \rho_{\text{Planckian,0}}\right)\frac{ne^2}{m^*}\frac{\hbar}{\alpha k_{\text{B}}}.$$

Here,

$$\rho_1 = A\frac{(k_{\text{B}}T_0)^2}{\hbar\varepsilon_F}\frac{m^*}{ne^2}.$$

Therefore,

$$T_c = \left\{A \frac{(k_B T_0)^2}{\hbar \varepsilon_F} + \alpha \frac{k_B T_{dec}}{\hbar} - A \frac{(k_B T_{dec})^2}{\hbar \varepsilon_F}\right\} \frac{\hbar}{\alpha k_B} .$$

Since $\varepsilon_F = k_B T_F$,

$$T_c = \frac{A{T_0}^2}{\alpha T_F} + T_{dec} - \frac{A}{\alpha} \frac{{T_{dec}}^2}{T_F}$$

$$= \frac{A{T_0}^2}{\alpha T_F} + \frac{\alpha}{2A} T_F - \frac{\alpha}{4A} T_F .$$

Hence,

$$T_c = \frac{A}{\alpha} \frac{{T_0}^2}{T_F} + \frac{\alpha}{4A} T_F .$$

In cuprates, $T_0 \ll T_F$ and $\frac{A}{\alpha} \frac{{T_0}^2}{T_F} \ll \frac{\alpha}{4A} T_F$,

$\therefore\ \boldsymbol{T_c \approx \frac{\alpha}{4A} T_F}$ : Uemura relation!

**Discussion S3: Linear fitting coefficient from experimental data of Uemura plot**

To obtain the coefficient, $\frac{\alpha}{4A}$ from Uemura relation in Eq. (7), we collect twenty experimental data set of $T_c$ and $T_F$ for cuprates (Table SI) from Uemura plot [39,40] and perform linear fitting for eleven data sets, excluding nine data points that deviated significantly from the linear relationship in the overdoped region (See Fig. S5). The coefficient is found to be 0.042 which is in agreement with experiment [41].

**Discussion S4: Analysis of Uemura relation by thermal decoupling**

In Fig. S3, the yellow line shows the resistivity of the Fermi liquid for the $CuO_2$ plane in the form of $T^2$. However, as we mentioned in the main text, the surface of YBCO is stabilized by the termination of the BaO plane, and the measured temperature corresponds to the effective temperature of the BaO plane ($T_{\mathrm{BaO}}$) which is higher than the effective temperature of the $CuO_2$ plane ($T_{\mathrm{CuO_2}}$) as shown in Fig. 4(b). Therefore, the measured temperatures are shifted as indicated by the green arrows and the resistivity becomes linear (green line) by the relation, $T_{\mathrm{BaO}} \propto {T_{\mathrm{CuO_2}}}^2$ (see section 3.3.1). The superconducting temperature of the $CuO_2$ plane is $T_0$ and the resistivity just before the superconducting transition is $\rho_1$ as shown in Fig. 5 and Fig. S3. If there are three kinds of high-$T_c$ superconductors (SC1, SC2, and SC3) of different $T_{\mathrm{dec}}$ ($T_{\mathrm{dec1}}$, $T_{\mathrm{dec2}}$, and $T_{\mathrm{dec3}}$), the measured critical temperature becomes $T_{c1}$, $T_{c2}$ and $T_{c3}$, respectively as shown in Fig. S3. The figure shows that $T_c$ increases from $T_{c1}$ to $T_{c3}$ as $T_{\mathrm{dec}}$ increases from $T_{\mathrm{dec1}}$ to $T_{\mathrm{dec3}}$. As mentioned in the main text, $T_{\mathrm{dec}}$ has a linear relationship with the $T_F$, $T_{\mathrm{dec}} = \frac{\alpha}{2A} T_F$ (Eq. (5)). Therefore, $T_{\mathrm{dec}}$ and $T_c$ increase as $T_F$ increases, as explained in Eq. (5) and (7).

**Discussion S5: Suppression of $T_c$ in overdoped cuprates: Analysis from thermal decoupling**

Usually, $T_F$ increases with the doping concentration in cuprates [39,40,75]. Thus, as the doping concentration increases, $T_F$ also increases, and consequently $T_c$ also increases, as shown in Fig. S5. However, as the doping concentration and $T_F$ increase further, $T_{\mathrm{c}}$ no longer increases but instead saturates and starts to decrease over the optimal doping concentration (see the green dashed curves in Fig. S5). We can consider the cuprate superconductor 3 (SC3) in Fig. S3 as the overdoped case, as the large $T_{\mathrm{dec3}}$ indicates a large $T_F$ and a high doping concentration. Overdoped superconductors typically enter the Fermi liquid regime as the temperature decreases, as indicated by $T_{\mathrm{FL}}$ (note that this is different from the Fermi temperature, $T_F$) in Fig. S3. Below $T_{\mathrm{FL}}$, the resistivity does not follow the linear-$T$ resistivity but follows quadratic temperature relation of the Fermi liquid. Thus, as indicated by the red dashed arrow in Fig. S3, the superconducting temperature is not expressed as $T_{\mathrm{c3}}$ but close to $T_{\mathrm{c2}}$, which is the reason why $T_{\mathrm{c}}$ is suppressed in the overdoped condition. The overdoped superconductor enters the Fermi liquid regime as the temperature decreases, because the thermal decoupling is maximized near $T_{\mathrm{FL}}$ and is suppressed below $T_{\mathrm{FL}}$. This explains the suppression and saturation of $T_{\mathrm{c}}$ on the increase of $T_F$ at higher $T_F$ of each cuprate, which is observed in the Uemura plot (see the green dashed curves in Fig. S5).

**Discussion S6: Least-square fitting of experimental data for Eq. (6)**

We perform least-square fitting for Eq. (6) from experimental data in Table SI. From the fitting for eleven data points in usual underdoped curates, we get $T_0$ = 13.9 K and $\frac{\alpha}{4A}$ = 0.043 as shown by the red line in Fig. S4. We also get $T_0$ = 13.5 K and $\frac{\alpha}{4A}$ = 0.041 as shown by the blue line in Fig. S4 for thirteen data points. From the fitting, we found that the superconducting temperature of $CuO_2$ plane, $T_0$ is about 14 K. We also found that $T_{\mathrm{c}}$ can be enhanced when $T_F$ is smaller than 200 K as shown in Fig. S4. It can explain the relation between superconductivity and flat band as discussed in section 4, since $T_F$ become extremely small in flat band

**Discussion S7: Calculation of isotope effect in cuprate superconductor from Eq. (6)**

In this Discussion, we show how to calculate the exponent **α** of isotope effect from Eq. (6) in the case of $YB_2C_3O_{6.6}$.

From Eq. (6) in main text, $T_c = \frac{A}{\alpha}\frac{{T_0}^2}{T_F} + \frac{\alpha}{4A}T_F$

YBCO ($YBa_2Cu_3O_{6.6}$) : $T_F = 1258.9$ K $\quad T_c = 54.9$ K

$T_0(^{16}O)$ = 14 K: superconducting temperature of $CuO_2$ plane from Discussion S6

The $CuO_2$ plane follows the Fermi liquid and BCS theory, so the exponent **α** of isotope effect for $CuO_2$ plane is 0.5. Therefore,

$\frac{T_0(^{18}O)}{T_0(^{16}O)} = \sqrt{\frac{16}{18}} \quad \therefore T_0(^{18}O)$ : 13.2 K

In the case of $^{16}O$, Eq. (6) becomes

$$54.9 = \frac{A}{\alpha}\frac{14^2}{1258.9} + \frac{\alpha}{4A} \times 1258.9$$

$x = \frac{\alpha}{A}$

$314.725x^2 - 54.9x + 0.1557 = 0$

$x = 0.17155 \quad = \frac{\alpha}{A}$

This is not affected by the isotope effect because electronic properties determine this.

In the case of $^{18}O$, Eq. (6) becomes

$T_c(^{18}\text{O})= \frac{1}{0.17155}\frac{13.2^2}{1258.9} + \frac{0.17155}{4} \times 1258.9 = 54.798$ K

$T_c(^{18}\text{O})$ decreases by 0.102 K compared to $T_c(^{16}\text{O})$.

Since $T_c{*}M^{\boldsymbol{\alpha}}$ is constant, the exponent $\boldsymbol{\alpha}$ of isotope effect**:**

$$\boldsymbol{\alpha} = \frac{\ln\frac{54.798}{54.9}}{\ln\frac{16}{18}} = 0.016$$

Using the same calculation, we obtain the exponent $\boldsymbol{\alpha}$ values for twenty cuprates (See Table SII). The average of $\boldsymbol{\alpha}$ values for twenty cuprates is 0.022 as mentioned in Section 3.3.4.

## Supplemental Figures

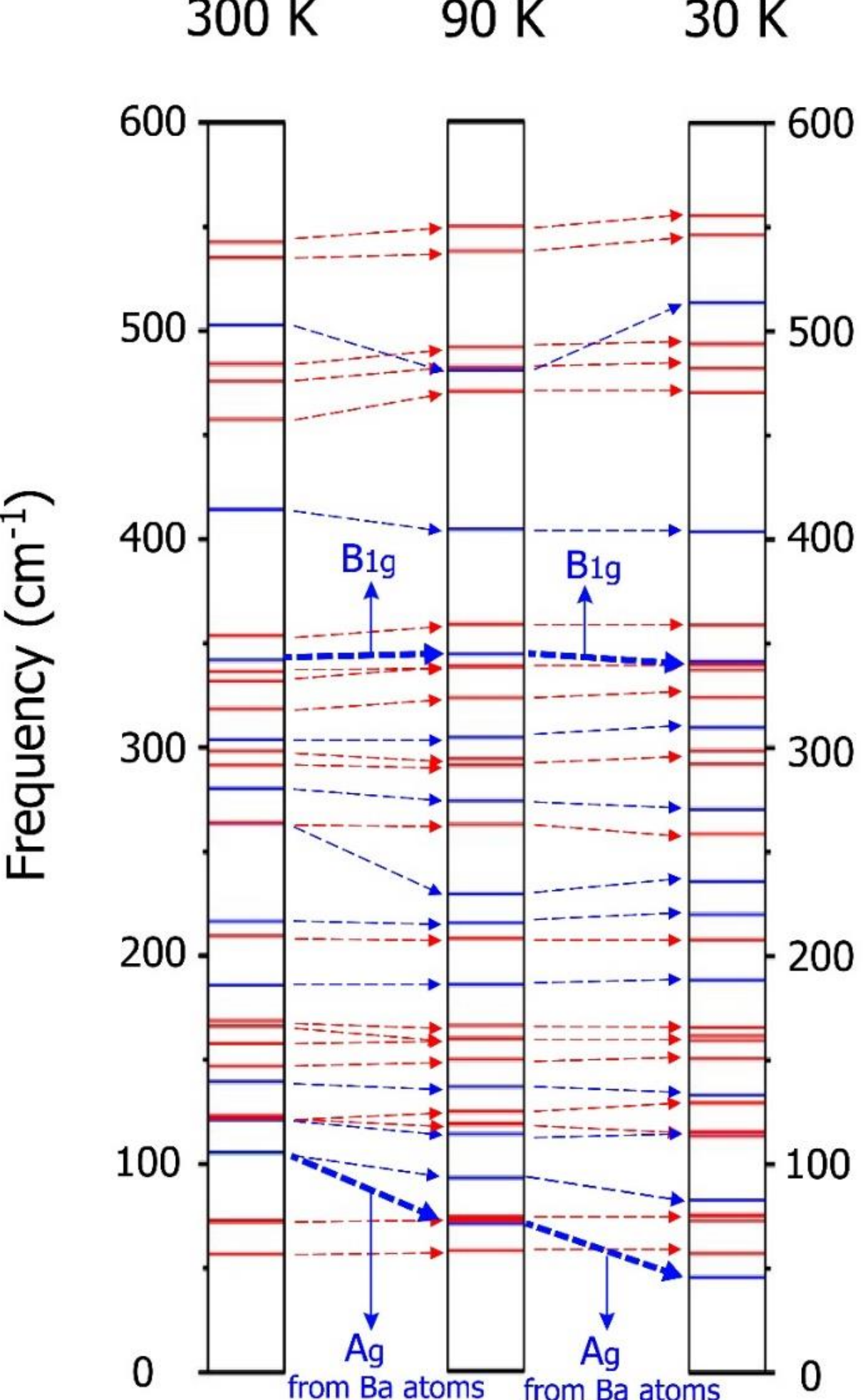

Fig. S1. Phonon frequencies at the Γ k-point at temperatures of 300, 90, and 30 K. The frequencies of the most out-of-plane phonons (shown in blue) decrease with the temperature, whereas the frequencies of the most in-plane phonons (shown in red) increase as the temperature decreases.

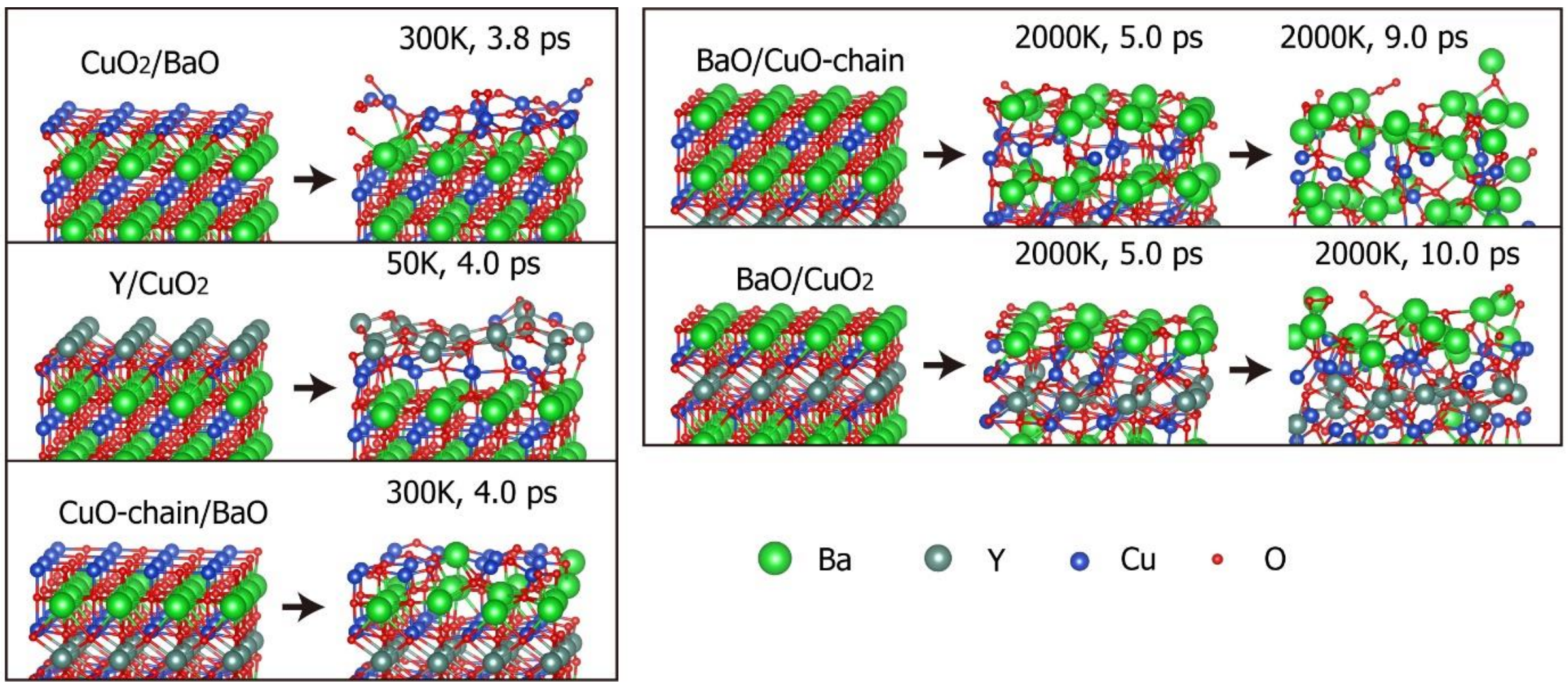

Fig. S2. Surface stability of YBCO from the *ab initio* molecular dynamics simulation. The surface structures in the figures are denoted as the top layer/ underlying layer. The $CuO_2$/BaO and Y/$CuO_2$ surface structures show the breaking of surface structures above the BaO layer in 4.0 ps at room temperature and 50 K, respectively. The CuO-chain/BaO surface structures show the breaking of the CuO-chain surface and the penetration of Ba atoms into the surface at room temperature. The simulations also show that the structures below the BaO layer sustain stability. The BaO/CuO-chain, and the BaO/$CuO_2$ surface structures reveal very stable surfaces below 2000 K. The BaO surfaces remain for 6 ps at 2000 K and then start to break. When the BaO surface is breaking, the underlying layers break together. From our *ab initio* molecular dynamics simulations, we find that the BaO surface acts as a protective surface layer of YBCO.

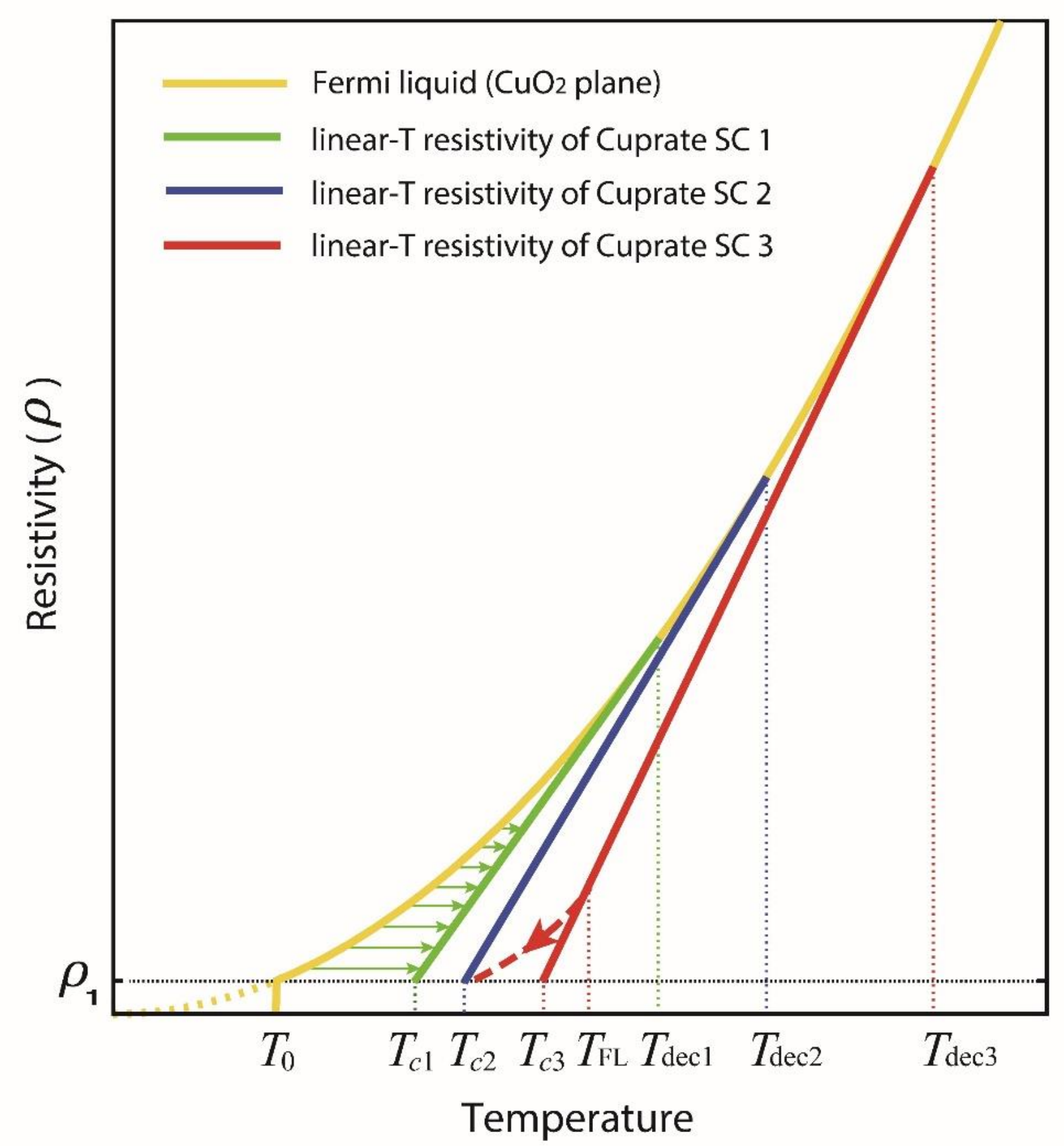

Fig. S3. Analysis of the Uemura relation and analysis of $T_c$ suppression in overdoped cuprates by thermal decoupling. $T_c$ increases from $T_{c1}$ to $T_{c3}$ as $T_{\mathrm{dec}}$ increases from $T_{\mathrm{dec1}}$ to $T_{\mathrm{dec3}}$. In the case of the overdoped superconductor in SC3 (red line), as the temperature decreases, it transitions to Fermi liquid at $T_{\mathrm{FL}}$causing $T_c$ decrease from $T_{c3}$ to $T_{c2}$.

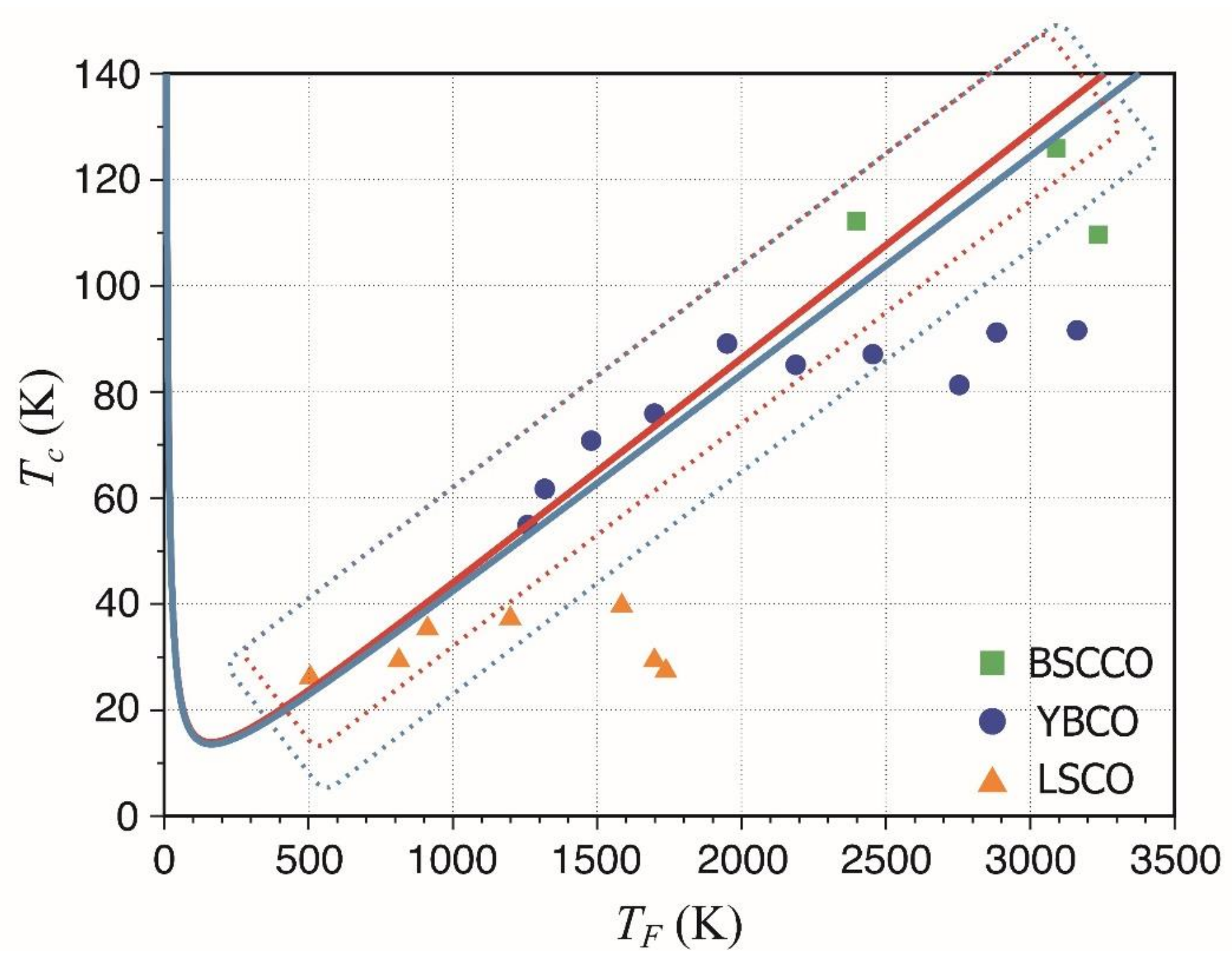

Fig. S4. Superconducting temperature ($T_0$) of $CuO_2$ plane in cuprates from the least-square fitting of experimental data. The fitting is performed for Eq. (6). Red and blue solid lines show the fitting for eleven data points in red dotted square and for thirteen data points in blue dotted square, respectively. The exclusion of seven data points from the fitting is due to the deviation from Uemura's linear relation by entering into Fermi liquids for overdoped cuprates as explained in Discussion S3 and Discussion S5

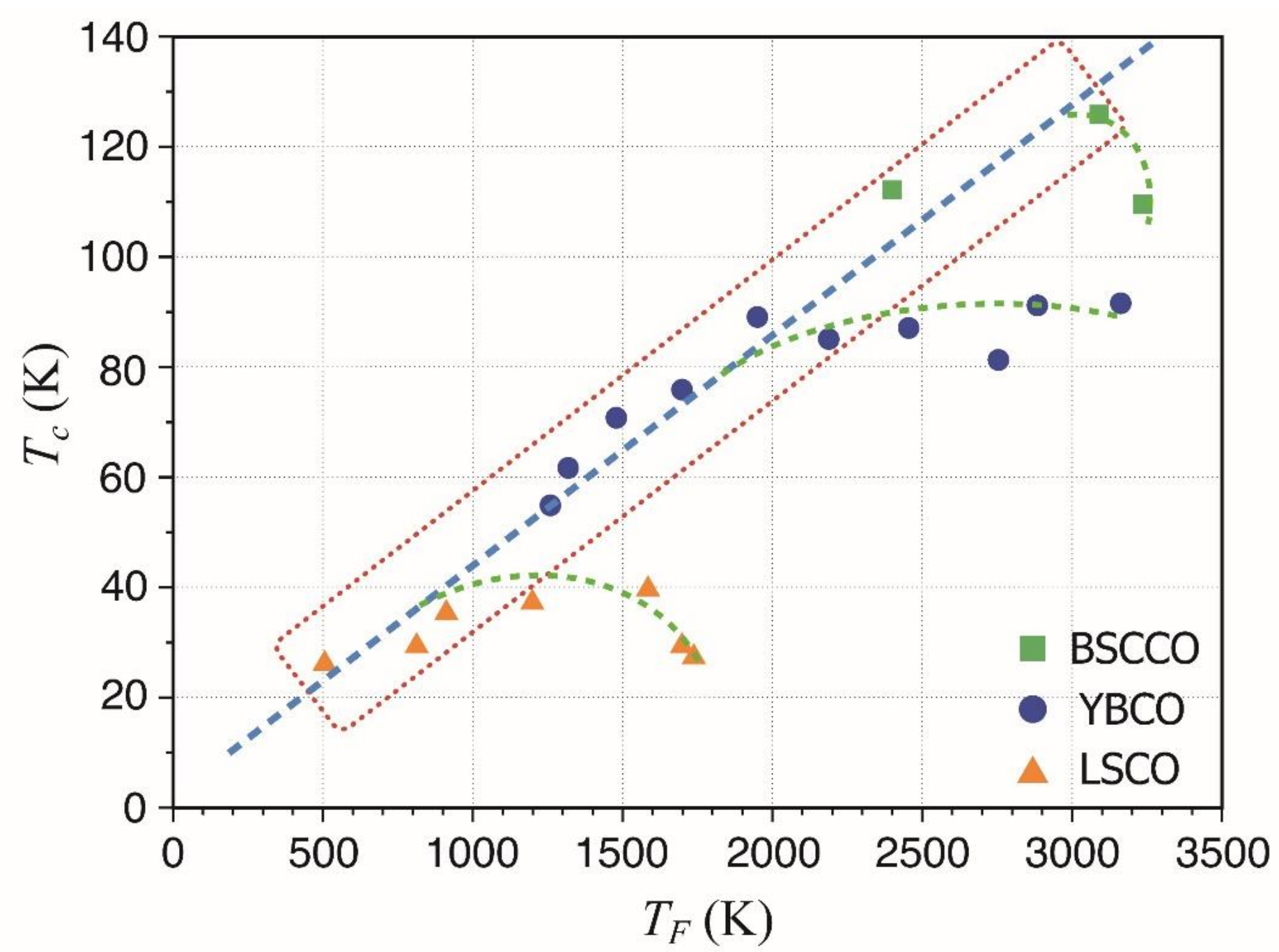

Fig. S5. Cuprates part of the Uemura plot in real scale. The blue dashed line, $y = 0.042x + 2.79$, is from the linear fitting of data inside the red dotted square. Green dashed curves show the saturation and suppression of $T_c$, which deviate from the fitting line for each cuprate as $T_F$ increases. The data points are from Uemura's papers [39,40] and Table SI.

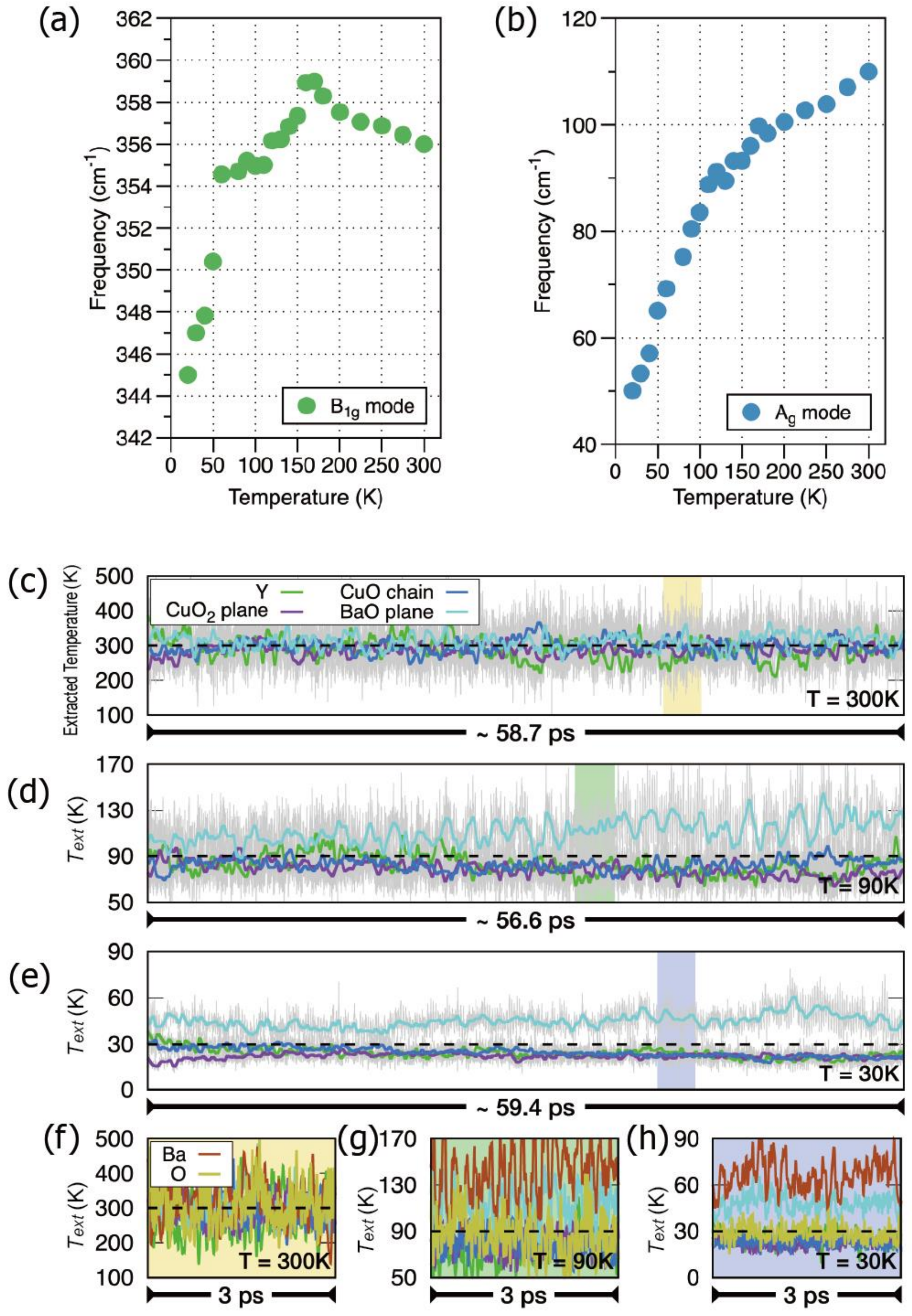

Fig. S6. $B_{1g}$ and $A_g$ phonon anomalies and thermal decoupling in the *ab initio* optimized structure (15.308 Å×15.652 Å×11.658 Å) of YBCO. The trends in phonon anomalies (a, b) and thermal decoupling (c–h) are similar to the results shown in Figs. 1 and 3, respectively. The phonon frequencies in the *ab initio*-optimized structure are overestimated compared to those in Fig. 1.

## Supplemental Tables

**Table SI.** Experimental data of $T_F$ and $T_c$ for cuprates from Uemura's plot [39, 40]. The numbers correspond to the data points in Fig. S4 and Fig. S5.

| Materials | Number | $T_F$ (K) | $T_c$ (K) |
|---|---|---|---|
| LSCO | 1 | 505.9 | 26.3 |
| LSCO | 2 | 812.8 | 29.5 |
| LSCO | 3 | 912.0 | 35.5 |
| LSCO | 4 | 1198.9 | 37.4 |
| LSCO | 5 | 1584.9 | 39.8 |
| LSCO | 6 | 1698.2 | 29.5 |
| LSCO | 7 | 1737.8 | 27.5 |
| YBCO | 1 | 1258.9 | 54.9 |
| YBCO | 2 | 1318.3 | 61.7 |
| YBCO | 3 | 1479.1 | 70.8 |
| YBCO | 4 | 1698.2 | 75.9 |
| YBCO | 5 | 1949.8 | 89.1 |
| YBCO | 6 | 2187.8 | 85.1 |
| YBCO | 7 | 2454.7 | 87.1 |
| YBCO | 8 | 2754.2 | 81.3 |
| YBCO | 9 | 2884.0 | 91.2 |
| YBCO | 10 | 3162.3 | 91.6 |
| BSCCO | 1 | 2398.8 | 112.2 |
| BSCCO | 2 | 3090.3 | 125.9 |
| BSCCO | 3 | 3235.9 | 109.6 |

**Table SII.** Exponent $\alpha$ values in the oxygen isotope effect of various cuprates calculated from Eq. (6),

$$T_c = \frac{A}{\alpha}\frac{{T_0}^2}{T_F} + \frac{\alpha}{4A}T_F$$

| Materials | Number | $T_F$ [a] (K) | $T_c(^{16}O)$ [a] (K) | $T_0(^{16}O)$ (K) | $T_0(^{18}O)$ (K) | $\alpha/A$ | $T_c(^{18}O)$ (K) | $\alpha$ |
|---|---|---|---|---|---|---|---|---|
| LSCO | 1 | 505.9 | 26.3 | 14 | 13.198 | 0.192 | 26.075 | 0.073 |
| LSCO | 2 | 812.8 | 29.5 | 14 | 13.198 | 0.136 | 29.303 | 0.057 |
| LSCO | 3 | 912.0 | 35.5 | 14 | 13.198 | 0.149 | 35.340 | 0.038 |
| LSCO | 4 | 1198.9 | 37.4 | 14 | 13.198 | 0.120 | 37.249 | 0.034 |
| LSCO | 5 | 1584.9 | 39.8 | 14 | 13.198 | 0.097 | 39.658 | 0.030 |
| LSCO | 6 | 1698.2 | 29.5 | 14 | 13.198 | 0.065 | 29.303 | 0.057 |
| LSCO | 7 | 1737.8 | 27.5 | 14 | 13.198 | 0.059 | 27.287 | 0.066 |
| YBCO | 1 | 1258.9 | 54.9 | 14 | 13.198 | 0.172 | 54.798 | 0.016 |
| YBCO | 2 | 1318.3 | 61.7 | 14 | 13.198 | 0.185 | 61.610 | 0.012 |
| YBCO | 3 | 1479.1 | 70.8 | 14 | 13.198 | 0.190 | 70.722 | 0.009 |
| YBCO | 4 | 1698.2 | 75.9 | 14 | 13.198 | 0.177 | 75.827 | 0.008 |
| YBCO | 5 | 1949.8 | 89.1 | 14 | 13.198 | 0.182 | 89.038 | 0.006 |
| YBCO | 6 | 2187.8 | 85.1 | 14 | 13.198 | 0.155 | 85.035 | 0.006 |
| YBCO | 7 | 2454.7 | 87.1 | 14 | 13.198 | 0.141 | 87.037 | 0.006 |
| YBCO | 8 | 2754.2 | 81.3 | 14 | 13.198 | 0.117 | 81.232 | 0.007 |
| YBCO | 9 | 2884.0 | 91.2 | 14 | 13.198 | 0.126 | 91.140 | 0.006 |
| YBCO | 10 | 3162.3 | 91.6 | 14 | 13.198 | 0.115 | 91.540 | 0.006 |
| BSCCO | 1 | 2398.8 | 112.2 | 14 | 13.198 | 0.186 | 112.151 | 0.004 |
| BSCCO | 2 | 3090.3 | 125.9 | 14 | 13.198 | 0.162 | 125.857 | 0.003 |
| BSCCO | 3 | 3235.9 | 109.6 | 14 | 13.198 | 0.135 | 109.550 | 0.004 |

[a] Reference [39, 40].